# Optimizing Stochastic Gradient Descent in Text Classification Based on Fine-Tuning Hyper-Parameters Approach.

## A Case Study on Automatic Classification of Global Terrorist Attacks.

Shadi Diab, Al-Quds Open University, Ramallah, Palestine. Email: shdiab@qou.edu

**Abstract--The objective of this research is to enhance performance of Stochastic Gradient Descent (SGD) algorithm in text classification. In our research, we proposed using SGD learning with Grid-Search approach to fine-tuning hyper-parameters in order to enhance the performance of SGD classification. We explored different settings for representation, transformation and weighting features from the summary description of terrorist attacks incidents obtained from the Global Terrorism Database as a pre-classification step, and validated SGD learning on Support Vector Machine (SVM), Logistic Regression and Perceptron classifiers by stratified 10-K-fold cross-validation to compare the performance of different classifiers embedded in SGD algorithm. The research concludes that using a grid-search to find the hyper-parameters optimize SGD classification, not in the pre-classification settings only, but also in the performance of the classifiers in terms of accuracy and execution time.**

**Keywords: Stochastic Gradient Descent; Classification of Terrorist attacks; SVM; Logistic Regression; Perceptron**

## I. Introduction And Literature Review

Natural language processing (NLP) is a collective term referring to the automatic computational processing of human languages [1], some of NLP applications are machine translation, speech recognition, information retrieval, clustering, and classification. The classification task aims to choose the correct class label for a given input, as long as each sample considered in isolation from all others, and a set of tags defined in advanced to perform classification [2]. Furthermore, many researchers studied different classification techniques in a database, data mining, and information retrieval communities [3]. Introducing a method for predicting possible future attacks is one of the central tasks of identifying the dynamic of terrorist attacks [4], which is a challenge.

_Shadi Diab_
_Al-Quds Open University_
_Kanan Building, Al Ersal St._
_PO Box: 1804_
_Ramallah – Palestine_
_Email: shdiab@qou.edu_

This research employs the Global Terrorism Database (GTD) built by the Study of Terrorism and Responses to Terrorism (START) consortium to investigate how to enhance the classification of terrorist attack incidents.

GTD has many details about worldwide-recorded terrorist incidents. Our research concerned about the summary variable of the attacks incidents that provides a brief narrative unstructured text. Moreover, we studied the performance metrics for different classifiers embedded in SGD optimization algorithm to improve the automatic classification of the terrorist attacks incidents. Different machine learning algorithms and techniques have been studied on data related to the terrorist attacks, [5] demonstrated a semantic network analyzer (SNAZ) to quantify sentiment in a collection of open source documents related to terrorist groups. Other researchers adopted the clustering-based anomaly detection algorithm to rank anomalous observations in GTD by using K-means clustering algorithm and validated their method by exploring the classification accuracy before and after detection of the outliers by using SVM, NB and LR [6].

[7] Classified terrorist groups based on the attacks patterns by analyzing textual descriptions of such attacks by using latent semantic indexing and clustering. [8]Analyzed different data preprocessing techniques for mining GTD to improve the classification of terrorist attacks in Iraq, the researcher concluded that data preprocessing and adding Global Positioning System coordinates could significantly reduce the classification error rate. [9] Presented the usage of clustering methods and association rule mining methods for discovering and representation of possible similarities in the data.

Other researchers introduced different visualization techniques to present various types of patterns and clarify the exploration of data related to terrorism. [10] Proposed a web-based interactive visual exploratory tool for GTD and [11] introduced a unified visualization environment to support the visualization of spatial-multivariate, spatiotemporal, temporal -multivariate, and spatiotemporal - multivariate. [12] Introduced an automatic capture of open source information from newspaper, twitter data related to the terrorist activities in India, and performed visual analytics on the Global Terrorism Database. [13] Proposed a visual analytical system that focuses on depicting the five W's (who, what, where, when, and why) in GTD.

Many other researchers studied SGD in Machine-Learning field for a long time. It received a significant attention and considered a very efficient approach to discriminative learning of linear classifiers under convex loss functions [14]. [15] Provided theoretically and experimental analysis of SGD over large-scale problems, in their results, they concluded that SGD performed well for large-scale problems, and second order stochastic gradient and averaged stochastic gradient are asymptotically efficient after a single pass on the training set. [16] Compared the standard SGD-based method in training time with a fuzzy kernel based, they utilized the fuzzy membership evaluation methods to transfer the distribution or geometry information carried in the data to the fuzzy memberships. [17] [18] Introduced Fuzzy and Support Vector Machine clustering by using SGD, [19] [20] [21] [22] investigated adapting of SGD and its variations in satellite image registration, face tracking, images registration, and images representation consequently.

In text classification based on SGD, [23] reviewed the effectiveness of different supervised and unsupervised learning approaches in text classification, and compared the performance of different SGD classifiers on four categories of newsgroup dataset. They proposed a plain stochastic gradient descent learning routine without mentioning the loss function, penalties and other parameters effect on the classification results. Moreover, their results did not reason out that K-Means had the highest accuracy score as clearly appeared in their designed experiment. [25] Described the Bangla document categorization using (SGD) classifier in comparison with Support Vector Machine (SVM) and Naïve Bayesian (NB) classifiers on collected newspapers articles. We consider this research is the most related research, but without mentioning the regularization of the dataset before learning the classifiers and its effect on the performance results. Furthermore, they did not mention and seemed they manually formed the parameters of the various components of the chain when structuring their experiment.

## II. The Classification Problem

Machine learning is a type of artificial intelligence (AI) that provides software to become more reliable in prediction [26]. The common routine in machine learning is to split the data into training and testing sets, the first set is used to learn the attributes of data and then evaluate such learned attributes on the unseen test set. In classification, we predict the correct class label for a given input value. We can define the classification problem mathematically as follows:

Let $D_n$ is a set of documents = $\{D_1, D_2, D_3 \ldots D_n\}$, where $n$ is the total number of the documents.

Let $C_n$ is a set of labels differentiates each class of the document (D), $C_k = \{C_1, C_2, C_3 \ldots C_k\}$, where k is the total number of labels.

The problem is to find the correct class ($C_k$) for each document ($D_n$)

To tackle this issue, we need an accurate classifier (f) to map $D_n$ to one of $C_k$ based on some criteria or selected features:

$$D_n \longrightarrow f \longrightarrow C_k$$

## III. Stochastic Gradient Descent (SGD) Learning

The learning process in machine learning is producing the function (f) by processing the samples of the training set; the function itself maps its input value $D_n$ to one of the classes $C_k$. In our research, we represented the text contents for each $D_k$ by extracting the numeric features vectors. Therefore, our feature vector extractor ($\varphi$) computes each vector feature for each input $\varphi (D_k) = \{\varphi1(D_k), \ldots, \varphi d(D_k) \}$, where $\varphi(D) \in \mathbf{R}d$ is a point in the dimensional space. Moreover, the parameter vector that specifies the contributions of each feature vector to the prediction process is $P = \{P_1 \ldots P_d\}$, where $P \in \mathbf{R}d$.

Consequently, we can mathematically calculate ($f$), by compiling both $\varphi$ (D) and $P$:

$$f = \varphi(D).P$$

The loss functions in Gradient Descent (GD) is the cost of inaccuracy of predictions, (GD) is an optimization straightforward algorithm aims to find the coefficient of (f) in a condition that minimizes the cost margin. It performs different coefficient values and the cost function estimates their cost through the predicted results for each sample of the training set. The aforementioned process occurs by comparing the prediction result with the actual value to choose the lowest price, and then the algorithm tries different coefficient value to look for lower one, then finally, it updates the coefficient by using a learning rate value to convert it on the next iteration. Such a calculation is very expensive and the cost is computed over the entire training dataset for each iteration. On the other hand, SGD updates the coefficient for each training sample and not at the end of the iteration over all samples of the training set.

In this research, we compared the performance of the SGD algorithm on different classifiers in term of accuracy and execution time, and by using different cost functions, various settings and different transformation of the raw data.

## IV. Research Methodology

After 1997, a team of GTD started creating the contents of the summary filed systematically by taking into consideration WH5 for each incident (when, where, who, what, how, and why) [27]. Next sub-sections describe the methodology steps in details.

### I. Collecting and Preparing Data

The GTD raw data have been collected from the National Consortium for the Study of Terrorism and Responses to Terrorism, hosted by the University of Maryland in the USA [28]. After ignoring the null values of the summary field, cleaning non-alphabetic characters, and removing all stop words, according to "Glasgow Information Retrieval Group" [29], we divided the data into different datasets, 70% for training and 30% for testing.

## II. Feature Extraction

We extracted the features by transforming the text to numerical features by the following steps:

- Converting the collection of text to numeric feature vectors by using n-grams representation (Vectorization), as a result of tokenization, counting, and normalization text.

- Weighting the count features by using TF-IDF (Term Frequency–Inverse Document Frequency) transformation.

TF used to assign the weight of the word (term (**t**)) depends on its occurrences in document (**d**) and donated by TF **t**, **d**. While the IDF is the number of documents containing (**t**) and donated by IDF **t**, where **t** is a term and **d** is a document. In our research, we computed TF-IDF vectors and normalized the results by using Euclidian normalization as in the following equations:

$$\textbf{TF-IDF (t, d) = TF (t, d) x IDF (t)}$$
$$\textbf{IDF (t)} = \log \frac{n_d}{df(d,t)} + 1$$
$$\textbf{Vnorm} = \frac{v}{||v||_2} = \frac{V}{\sqrt{V1^2 + V2^2 + \cdots + Vn^2}}$$

## III. Experiment and Classification Design

Our experiment aims to compare the performance of SVM, Logistic Regression and Perceptron classifiers in the following three different situations:

- Applying the default parameters.
- Applying SGD-learning and Non-SGD-learning.
- Applying SGD-learning with the hyper-parameters approach.

## IV. Validation and Evaluation

We evaluated the performance of each classifier on different situations and settings. However, In order to avoid the overfitting problem, while the knowledge about the testing set cannot fit perfectly into the model and evaluation metrics no longer report on performance [30], we created a validation set from the training set, and then our evaluation applied to the unseen dataset. We evaluated the final performance by creating a confusion matrix to find out the values of True Positive (TP), True Negatives (TN), False Positives (FP), False Negatives (FN) values to compute the following performance metrics:

- Accuracy: To calculate the percentage of samples in the test set that the classifier correctly labeled, measured by calculating TP+TN/TP+FP+FN+TN [2]

- Precision: Indicates how many incidents were relevant, and measured by TP/ (TP+FP) [31].

- Recall: Indicates how many of the relevant items were identified, and measured by TP/ (TP+FN) [31].

- F1 combines the precision and recall to give a single score. It is the harmonic mean of the precision and recall and calculated by (2 × Precision × Recall)/ (Precision Recall) [31].

- Macro-Average: To calculate the harmonic mean of precision, recall, support and other metrics for all classes.

## V. Data preprocessing

GTD database has 170,350 different attack incidents classified into nine attacks types: (Assassination, Armed Assault, Unarmed Assault, Hijacking, Kidnapping, Barricade Incident, Bombing/Explosion, Facility/Infrastructure, and Unknown Attack Type). Moreover, each attack type is coded into numeric values from 1 to 9 respectively, we extracted the summary field for each category attack's type. After cleaning the dataset by removing all null summary fields and removing stop words, the total number of the extracted summaries became 102,669 incidents distributed as shown in figure 1.

*Figure 1:  Distribution of attack types from 1997 to 2016*

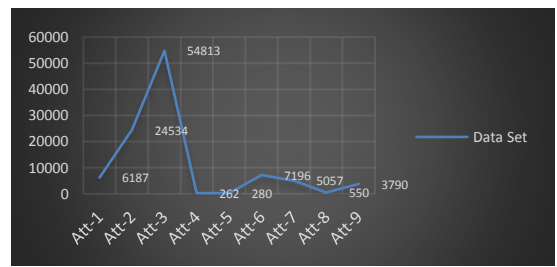

We divided the data into 70% for training purpose, and 30% for testing. Table 1 clarifies the distribution of attacks types for both training and testing sets.

*Table 1: Distribution of attacks types in training and testing sets*

| Attack Type | Training Set | Testing Set |
|---|---|---|
| **Attack Type-1** | 4372 | 1815 |
| **Attack Type-2** | 17182 | 7352 |
| **Attack Type-3** | 38335 | 16478 |
| **Attack Type-4** | 192 | 70 |
| **Attack Type-5** | 188 | 92 |
| **Attack Type-6** | 5075 | 2121 |
| **Attack Type-7** | 3498 | 1559 |
| **Attack Type-8** | 356 | 194 |
| **Attack Type-9** | 2670 | 1120 |
| **Totals** | **71868** | **30801** |

As shown in figure 1 and table 1, (53.3%) of the dataset samples belong to category 3 (Att-3). Such skewed data may cause accuracy paradox and bias against some classes when evaluating the performance. There are several methods used to tackle this problem, in our experiment we applied SMOTE (Synthetic Minority Over-sampling Technique) to over-sample the minority classes of the dataset through generating synthetic examples rather than over-sampling with replacement [32]. Moreover, Stratified K-Folds cross-validation used in the validation process where the folds made by maintaining the percentage of samples for each class.

We addressed the effect of such solutions when comparing the performance metrics when applying SMOTE method and without applying the method to resolve the data-skewing problem, as clarified in figure 2 and 3, despite the accuracy of the classifier is greater without balancing classes, It has unpredicted samples in class 4. On the other hand, the accuracy of the classifier after applying SMOTE become less score without any ill precision or ill recall values in class 4. In conclusion, balancing categories as a step before training the classifier will lead to reliable and more accurate classification. In our experiments, SMOTE has been applied in the final evaluation and Stratified K-Folds cross-validation in the validation step.

*Figure 2: Example - Accuracy before applying SMOTE*

```
Accuracy of SGD Linear SVM before applying SMOTE:
0.858348754911
              precision    recall  f1-score   support

        cat0    0.66752   0.14380   0.23663      1815
        cat1    0.73197   0.90860   0.81078      7352
        cat2    0.93302   0.97554   0.95381     16478
        cat3    0.87500   0.30000   0.44681        70
        cat4    0.00000   0.00000   0.00000        92
        cat5    0.87626   0.94154   0.90773      2121
        cat6    0.82096   0.75882   0.78867      1559
        cat7    1.00000   0.10309   0.18692       194
        cat8    0.69072   0.17946   0.28490      1120

avg / total    0.84850   0.85835   0.83272     30801
```

*Figure 3: Example - Accuracy after applying SMOTE*

```
Accuracy of SGD Linear SVM after applaying SMOTE:
0.848024414792
              precision    recall  f1-score   support

        cat0    0.67568   0.08264   0.14728      1815
        cat1    0.71950   0.90574   0.80195      7352
        cat2    0.92351   0.97530   0.94870     16478
        cat3    0.87500   0.20000   0.32558        70
        cat4    0.24138   0.68478   0.35694        92
        cat5    0.89246   0.91561   0.90389      2121
        cat6    0.85315   0.70430   0.77161      1559
        cat7    1.00000   0.06701   0.12560       194
        cat8    0.65089   0.09821   0.17067      1120

avg / total    0.84293   0.84802   0.81774     30801
```

## VI. Fine-Tuning Hyper-Parameters

Hyper-Parameters effects on the performance of the classification model when having a high number of options for the selected parameters. In our main experiment, we investigated the effects of the hyper-parameters when comparing the performance for the same classifiers and by using SGD learning. Extracting the best values of the parameters to optimize the performance metrics is a challenge. In our optimization experiment, we implemented a grid-search method to compute and compare the performance metrics for the same classifiers and by using different parameters with different values, to get the best combination of parameters that optimize the performance of the classification task.

We created 50% development set to find out the best scores for each selected parameter. In our grid-search approach, we computed the accuracy and execution time of SVM, Logistic Regression and Perceptron classifiers respectively. The mean (μ) and the standard deviation (σ) returned by using different permutations according to ngram_rage, normalization, use IDF, use smooth IDF, penalty, and alpha parameter. Table 2 clarifies the best permutation of the parameters that achieved the best mean (μ) and (σ).

*Table 2: Permutations of parameters achieved best metrics*

| Classifiers | μ | σ | Parameters |
|---|---|---|---|
| SVM | 0.80 | (+/-0.12) | (1,2),'l2',True,True,'l2',1e-05 |
| Logistic Reg. | 0.83 | (+/-0.10) | {(1, 1),'l2',True,False,'l2', 1e-05} |
| Perceptron | 0.77 | (+/-0.13) | (1, 2),'l2',True,True,'l2',1e-04 |

## VII. Experiment and Validation

In our research, we calculated the mean of accuracy and the standard deviation for ten validation folds, to find out the performance metrics of different linear classifiers by using SGD learning and Non-SGD learning and finally explored optimizing SGD classification for the same classifier by applying our approach.

We implemented our work by using Python programming language and by different python models, the open source-data analysis library (Pandas) model [33] scipy.sparse data structure [34], Scikit-learn module in python [35] and by using 10 K-Fold Cross-Validation. We tokenized each document, removed stop words, structured a dictionary of features, and transformed documents to feature vectors, normalized and weighted the count features by TF-IDF.

We implemented SVM, Logistic Regression and Perceptron classifiers by SGD learning and by using the default values of the classifiers. We explored the effects of the following data processing parameters:

1. NGRAM_RAGE: determine the range of sequence of n items of the text, such as word, pairs of words.
2. NORM: the normalization method, two methods investigated, the least absolute deviations L1 (LAD), and the least square method L2.
3. USE_IDF: to enable or disable inverse-document-frequency when scoring the count features.
4. SMOOTH_IDF: to enable or disable using of the smoothing version of IDF by adding (1) as a numeric value to the formula of IDF to avoid division by zero or in the situation of the term given zero while appearing in all documents.
5. PENALTY: the aka regularization, two options investigated L1 and L2.
6. ALPHA: a constant used to compute the step size and the learning rate.

We measured the accuracy and the execution time in different settings, on SGD learning, Non-SGD learning and SGD learning with the best hyper-parameters. The training dataset divided into 10 different smaller datasets to validate the performance of each classifier under each situation, in our classification design stage, in which each classifier is trained by using K-1 of the folds as a new training set and validated by using the remaining fold. We measured two terms for validation, the mean of the accuracy value for each classifier and the execution time on Intel Core i7-5600U CPU @ 2.60 GHz (4 CPUs), and 8 GB RAM.

Table 3 summarizes the performance metrics of stratified 10 K-Folds Cross-Validation by using SGD learning and the default parameters of the classifiers.

*Table 3: 10 K-folds validation with SGD learning*

| SGD Classifier | Accuracy | | Time / Sec. |
|---|---|---|---|
| | μ | σ | |
| SVM | 0.85571 | (+/- 0.01193) | 7.66101 |
| Logistic Reg. | 0.83709 | (+/- 0.01121) | 11.64996 |
| Perceptron | 0.85333 | (+/- 0.02513) | 7.68477 |

On the other hand, we validated the same classifiers by using the same validation settings, but without using SGD learning for each classifier, table 4 clarifies the performance metrics without using SGD learning.

*Table 4:10 K-folds validation without SGD learning*

| Classifier | Accuracy | | Time / Sec. |
|---|---|---|---|
| | μ | σ | |
| SVM | 0.87829 | (+/- 0.01544) | 31.33962 |
| Logistic Reg. | 0.86872 | (+/- 0.01556) | 89.45771 |
| Perceptron | 0.85306 | (+/- 0.02237) | 7.98505 |

Table 5 clarifies the validation results of SGD learning with best values obtained from finding the hyper-parameters approach

*Table 5:10 K-folds validation on SGD with hyper-parameters*

| Classifier | Accuracy | | Time / Sec. |
|---|---|---|---|
| | μ | σ | |
| SVM | 0.87946 | (+/- 0.01380) | 10.62 |
| Logistic Reg. | 0.87192 | (+/- 0.01122) | 8.07 |
| Perceptron | 0.87378 | (+/- 0.01543) | 10.59 |

Considering the result of our validation experiment, it becomes clear that using SGD learning with hyper-parameters improves the accuracy and the execution time of SGD on (SVM, Logistic Regression, and Perceptron) classifiers.

## VIII.    Evaluation

We evaluated our approach by using the unseen testing dataset on SVM, Logistic Regression, and Perceptron, by comparing the two situations, when using the default values of SGD learning, and using the best hyper-parameters obtained through our approach. The result of evaluation of the unseen 30% testing set before and after applying the best combination of parameters clarified in tables 6 and 7, where improvement on the accuracy of the classifiers taken place after applying best combinations of hyper-parameters.

*Table 6: the performance metrics by the proposed approach*

| Classifier | Performance Metrics | | | | Time / Sec |
|---|---|---|---|---|---|
| | Accuracy | Precision | Recall | f1 | |
| SVM | 0.885 | 0.879 | 0.885 | 0.878 | 30.04 |
| LR. | 0.870 | 0.865 | 0.870 | 0.861 | 26.81 |
| Perceptron | 0.856 | 0.486 | 0.432 | 0.457 | 25.44 |

*Table 7: The performance metrics by the default values*

| Classifier | Performance Metrics | | | | Time / Sec |
|---|---|---|---|---|---|
| | Accuracy | Precision | Recall | f1 | |
| SVM | 0.848 | 0.844 | 0.848 | 0.819 | 26.25 |
| LR. | 0.820 | 0.818 | 0.820 | 0.792 | 27.64 |
| Perceptron | 0.852 | 0.854 | 0.852 | 0.852 | 25.53 |

## IX.    Conclusion

In this research, we proposed Stochastic Gradient Descent learning to optimize text classification on Global Terrorism Database based on finding the best hyper-parameters. We applied SVM, Logistic Regression and Perceptron classifiers to classify the terrorist attacks incidents obtained from GTD, to measure the effects of finding hyper-parameters to optimize the performance of such classifiers.   In our experiment and the validations process, we explored different experiments to compare the performance of the classifiers without SGD learning, with SGD learning and with SGD learning including hyper-parameter discovery. Our research concludes that SGD learning optimizes the accuracy of the selected classifiers, and reduces the execution time for Logistic Regression and Perceptron classifiers.   Furthermore, applying a grid - search approach to find the hyper-parameter to justify the classifiers before training the classifiers will increase the accuracy of the classifiers, and enhance the whole classification technique.

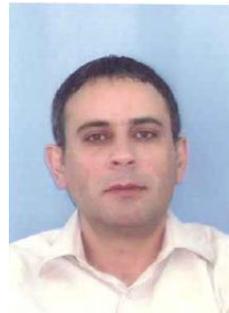

**Shadi Diab:** Over the past 17 years, Mr. Shadi Diab has been an information technology professional serving the education sector in Palestine, he has gained broad experience in different domains of information technology including education, training, management, certifications & testing systems. Currently, he is head of accreditation & internet based testing unit at the ICT center of Al-Quds Open University. Mr. Diab holds MSc degree in Computer Science and has several accredited certifications not only in information technology, but also in education, and management. mailto:shdiab@qou.edu , shdiab2@gmail.com